\documentclass[12pt]{article}
\usepackage[dvips]{graphicx}
\usepackage{times}
\usepackage{epsfig}
\usepackage{amssymb}
\usepackage{amsmath}
\usepackage{multirow}

\begin{document}
\begin{center}
{\large\bf   
Investigation of the electromagnetically induced transparency in era of cosmological hydrogen recombination\bigskip\\} 
${}^{a,b}$D. Solovyev, ${}^{c,d}$V. K. Dubrovich and ${}^e$G. Plunien
\medskip\\    
$^a$ V. A. Fock Institute of Physics, St. Petersburg
State University, Petrodvorets, Oulianovskaya 1, 198504,
St. Petersburg, Russia\\
$^b$ Max-Planck-Institut f\"{u}r Physik Komplexer Systeme, 
N\"othnitzer Str. 38, D-01187, Germany\\
$^c$ St. Petersburg Branch of Special Astrophysical Observatory, Russian Academy of Sciences, 196140, St. Petersburg, Russia \\
$^d$ Nizhny Novgorod State Technical University n. a. R. E Alekseev, LCN, GSP-41, N. Novgorod, Minin str., 24, 603950 \\
$^e$ Institut f\"{u}r Theoretische Physik, Technische Universit\"{a}t Dresden, Mommsenstrasse 13, D-01062 Dresden, Germany \bigskip\\
E-mail: solovyev.d@gmail.com
\end{center}

\begin{abstract}
Investigation of the cosmic microwave background formation processes is one of the most actual problem at present time. In this paper we analyze the response of the hydrogen atom to the external photon fields. Field characteristics are defined via conditions corresponding to the recombination era of universe. Approximation of three-level atom is used to describe the "atom - fields" interaction. It is found that the phenomena of the electromagnetically induced transparancy (EIT) takes place in this case. Consideration of EIT phenomena makes it necessary to update astrophysical description of the processes of the cosmic microwave background formation and, in particular, Sobolev escape probability. Additional terms to the optical depth entering in the Sobolev escape probability are found to contribute on the level about $1\%$.
\end{abstract}

\section{Introduction}
In view of recent success in the theoretical description and experimental observations of the cosmic microwave background (CMB) a detailed analysis of all the processes occurring in cosmological recombination era is required. For a precise description of CMB in astrophysical investigations the different photon-emission and absorption, photon-electron scattering and etc. processes should be included. The spectral characteristics of the emitted radiation due to different phenomena, the polarization state and the cross-section are particularly relevant from the astrophysical point of view. In standard calculations for such kind of tasks the quantum mechanical approach for the isolated atom is applied.

However, the consideration of the "atom-fields" interaction becomes important in case of astrophysical experiments with accuracy of about $1\%$ \cite{Hin}, \cite{Page} and with 
the expectation of even increasing the accuracy up to the level of $\sim 0.1\%$. 
Atom's interaction with an external fields can lead to effects like population inversion, steady-state solution, line strengths, cross sections, susceptibility and polarization and etc \cite{Bonch}-\cite{Li-Xin}. Apart from other powerful methods (Green function approach, for example) the statistical operator theory can be applied for the study of the "atom-field" systems. Application of the density matrix formalism in three-level approximation seems to be more simple and appropriate in this case. The clear description of the density matrix theory and its applications, such as spontaneous emission, line broadening (power broadening and saturation, collision line broadening, Doppler broadening and Voigt profile), can be found, for example, in \cite{Weiner}.

The radiation transfer theory that is usually applied in the research of cosmic microwave background was suggested in \cite{Zeld}, \cite{Peebles}.  In particular, it was established that the $2s\leftrightarrow 1s$ transition is able to substantially control the dynamics of cosmological hydrogen recombination. Moreover, distortions of the order $10^{-6}$ were predicted \cite{Zeld}. Recently, the radiation transfer theory for the recombination era of early universe was intensively re-examined due to precise observations of the CMB \cite{Seager}, \cite{Seager2}. In \cite{GD}, \cite{Dubr} corrections to the ionization history were found that exceed the percent-level. In order to achieve this accuracy the multi-photon decays in atoms involving set of states should be included in evaluation of CMB formation \cite{Chluba1}-\cite{Rubin2}. As a rule the two-photon decays of the excited states are considered for the evaluation of CMB. Two-photon emission processes were evaluated accurately in \cite{LSP}-\cite{SDVLP}. In our recent works \cite{PRA}, \cite{Can} three- and four-photon transitions with separation out of the two-photon links were considered from astrophysical point of view. In \cite{extf1}, \cite{extf2} we considered one- and two-photon transitions in external electric fields. Such modifications should have a strong impact to the determination of the key cosmological parameters \cite{Lewis}.

Generally, the absorption coefficient calculated per atom is used for the study of the radiative transfer in spectral lines. The influence of powerful high-frequency electromagnetic radiation on the absorption coefficient in the low-frequency line in a three-level $\Lambda$ atom was considered in \cite{Kaplanov}. In this paper we consider the other kind of multi-photon process, namely, the electromagnetically induced transparency (EIT) phenomena. The nature  of EIT phenomena 
can be examined by evaluation of the response of the multilevel system to the presence of an external radiation field. Electromagnetically induced transparency leads to a significant modification of the absorption profile of the system. Description of EIT phenomena for a three-level ladder system interacting with two near-resonant monochromatic fields can be found, for example, in \cite{Whitley}-\cite{Wiel}. A complete and detailed treatment of a three-level ladder atom, absorption and emission spectra, as the transient and steady-state response of the $\Xi$-atom are given in \cite{Whitley}. We study the response of the three-level ladder system on the external fields originating from photons emitted during the recombination and evaluate the absorption coefficient, which we apply to radiation transfer theory. The physics of an atom interacting with photon fields can be understood on the basis of the "interfering-pathways" description, which corresponds to multi-photon process defined in term of a power series expansion over fields amplitudes (see e.g. Refs. \cite{Whitley}, \cite{Gea-Ban} and \cite{Wiel}). 

Our paper is organized as follows: In section 2 we shall briefly review essentials of the density-matrix approach when deriving the density matrices element $\rho_{21}$ and presenting its  series expansion. In section 3 we shall define the absorption coefficient and illustrate its application to astrophysics on the basis of the \cite{Seager2}. In the subsequent sections we shall provide numerical calculations together with a discussion of the results. 
We conclude the paper with a summary.  

\section{Three-level $\Xi$ atom and density matrix}

In this section we employ the three-level ladder scheme for the description of the hydrogen atom. We assume that the hydrogen atoms formed during the recombination epoch in earlier universe reach their ground states via emission of photons of all the spectral lines corresponding to atomic continuum-bound and bound-bound transitions. All the emitted photons generate the 
(coherent) external field environment, which feats back onto the hydrogen atom. Investigation of this self-consistent scenario under the conditions of cosmic expansion is the aim of the present paper. 
We confine our consideration to the spontaneous emission rates. Collisional excitation and ionization can be omitted because at the relevant temperatures and densities they are negligible for a three-level hydrogen atom \cite{Seager2}.

To identify the possible effect we discuss atoms subjected to an external field 
which mainly consists of two adjacent spectral lines, namely the $Ly_{\alpha}$ and $H_{\alpha}$ lines, with the initial condition corresponding to full population of the atomic ground state.   Then we use the standard density matrix formalism which can be found, for example, in 
Ref. \cite{Wiel}.

The generic three-level ladder system is depicted in Fig. 1. 
Solving the density-matrix equations employing the steady state and the rotating-wave approximations yields the following set of equations for the relevant density-matrix elements \cite{Wiel}:
\begin{eqnarray}
\label{1}
\rho_{21}=\frac{i/2[\Omega_{\alpha}(\rho_{22}-\rho_{11})-\Omega_{\beta}^*\rho_{31}]}{\gamma_{21}-i\Delta_{\alpha}}\, ,
\nonumber
\\
\nonumber
\rho_{32}=\frac{i/2[\Omega_{\beta}(\rho_{33}-\rho_{22})+\Omega_{\alpha}^*\rho_{31}]}{\gamma_{32}-i\Delta_{\beta}}\, ,
\\
\rho_{31}=\frac{i/2[\Omega_{\alpha}\rho_{32}-\Omega_{\beta}\rho_{21}]}{\gamma_{31}-i(\Delta_{\alpha}+\Delta_{\beta})}\, ,
\\
\nonumber
\rho_{22}=\frac{i}{2\Gamma_2}(\Omega^*_{\alpha}\rho_{21}-\Omega_{\alpha}\rho_{12})\, ,
\\
\nonumber
\rho_{33}=\frac{i}{2\Gamma_3}(\Omega^*_{\beta}\rho_{32}-\Omega_{\beta}\rho_{23})\, .
\end{eqnarray}
The levels of system are specified with the following hydrogenic states 
$|1\rangle =|1s\rangle$, $|2\rangle=|2p\rangle$ and $|3\rangle=|3s\rangle$,
The frequencies $\omega_{21}$ and $\omega_{32}$ correspond to the transitions 
$|1\rangle\rightarrow |2\rangle$ and $|2\rangle\rightarrow |3\rangle$, respectively. 
The system is driven by a "probe" field with amplitude $E_{\alpha}$ at frequency $\omega_{\alpha}$ and by the "control" field with amplitude $E_{\beta}$ and frequency $\omega_{\beta}$.  $\Delta_{\alpha}=\omega_{\alpha}-\omega_{21}$, $\Delta_{\beta}=\omega_{\beta}-\omega_{32}$ 
define the corresponding de-tunings, $\Omega_{\beta}=2d_{32}E_{\beta}$ and $\Omega_{\alpha}=2d_{21}E_{\alpha}$ are the Rabi frequencies, which depend on the atomic dipole-matrix element $d_{ij}$. All these expressions are given in atomic units. 
Neglecting collisional de-phasing effects 
the decay rate is given by $\gamma_{ij}=(\Gamma_i+\Gamma_j)/2$, where $\Gamma_i$ is the natural decay rate of the level $|i>$.

In the limit of a weak probe field and $\rho_{11}\approx 1$, $\rho_{22}\approx\rho_{33}\approx 0$ ( full population of the ground state of the atom) the solution of the equations (\ref{1}) for the $\rho_{21}$ to first order in the probe field and to all orders in the control field was found in \cite{Gea-Ban}, \cite{Wiel}. But the total solution of Eq. (\ref{1}) for $\rho_{21}$ is
\begin{eqnarray}
\label{2}
\rho_{21}=\frac{-i \Omega_{\alpha}/2 \left(\frac{\Omega_{\alpha}^2}{4} + A\right)}
{\frac{\Omega_{\beta}^2}{4} \left(
      \frac{\Omega_{\alpha}^2}{4\Gamma_2} - 
      B\right) - (-
       \frac{\Omega_{\alpha}^2}{4\Gamma_2} + \gamma_{21} - 
      i \Delta_1) (\frac{\Omega_{\alpha}^2}{4} + A)}\, ,
\\
\nonumber
A=B \left(\gamma_{31} - i (\Delta_1 + \Delta_2)\right)\, ,
\\
\nonumber
B=\frac{\Omega_{\beta}^2}{4\Gamma_3} + \gamma_{32} - i \Delta_2\, ,
\end{eqnarray}
which can be easily transformed in the approximation of weak fields to the expression (2) in \cite{Wiel}.

The underlying physics of the systems response on external fields can be understood 
considering the power series expansion of the solution (\ref{2}) for the matrix element $\rho_{21}$ 
with respect to the variables $\Omega_{\alpha}$ and $\Omega_{\beta}$. 
Authors \cite{Wiel} derived a series that contains terms up to third order in $\Omega_{\alpha}$ and $\Omega_{\beta}$ at zero detuning. For our purposes it is important to keep non-zero detuning and, as before, in the approximation of weak field expression (\ref{2}) 
will be expanded into a power series with respect $\Omega_{\alpha}$ and $\Omega_{\beta}$.

The resulting expression for the $\rho_{21}$ simplifies significantly in case of exact two-photon resonance, i.e. when the frequencies of two external fields coincide exactly with the transition frequency $\omega_{31}=E_3-E_1$. In this case the equality $\Delta_{\alpha}+\Delta_{\beta}=0$ 
holds and the series expansion looks like
\begin{eqnarray}
\label{5}
\rho_{21}=
\frac{i\Omega_{\alpha}/2}{\gamma_{21}-i\Delta_{\alpha}}\left[1-\frac{\Omega_{\beta}^2/4}{\gamma_{31}(\gamma_{21}-i\Delta_{\alpha})}+
\frac{\Omega_{\alpha}^2/4}{\Gamma_2(\gamma_{21}-i\Delta_{\alpha})}+
\right.
\\
\nonumber
\left.
\frac{\Omega_{\alpha}^2/4\cdot\Omega_{\beta}^2/4}{(\gamma_{21}-i\Delta_{\alpha})^2(\gamma_{32}-i\Delta_{\beta})}\frac{(\Gamma_2+\gamma_{31})(\gamma_{21}-i\Delta_{\alpha})-2\gamma_{31}(\gamma_{32}-i\Delta_{\beta})}{\Gamma_2\gamma_{31}^2}+...
\right],
\end{eqnarray}
where the dots imply the higher-order terms in $\Omega_{\alpha}$, 
$\Omega_{\beta}$ and product $\Omega_{\alpha}\cdot\Omega_{\beta}$. The series expansion is done under the conditions $\Omega_{\beta}/\gamma_{ij}\ll 1$ and $\Omega_{\alpha}/\gamma_{ij}\ll 1$. 

As it was established in \cite{Wiel} the common pre-factor in (\ref{5}) corresponds to the one-photon absorption processes, while the squared terms are associated with the two-photon absorption and subsequent emission processes. The products $\Omega^2_{\alpha}\cdot\Omega^2_{\beta}$ represent the "interfering-pathways" terms, see Fig. 2. Thus the matrix element $\rho_{21}$ describes the multi-photon processes of the coherent "atom-field" interaction. 
For a detailed analysis of Eq. (\ref{5}) we refer to \cite{Wiel}.

\section{Absorption coefficient and Sobolev escape probability.}

The theory of radiation transfer for multilevel atoms 
utilizing the concept of the Sobolev escape probability 
has been described in \cite{Seager2}. 
With the method of escape probability a simple solution to the radiative transfer problem for all bound-bound transitions can be found. The Sobolev escape probability $p_{ij}$ 
($j$ refers to the upper level and $i$ to the lower level of a multilevel atom)
is the probability that photons associated with this transition will "escape" without being further scattered or absorbed. If $p_{ij}=1$, the photons produced in the line transition 
can escape to infinity - they do not give rise to distortions of the radiation field. If $p_{ij}=0$, the  photons can not escape to infinity; all of them get re-absorbed, and the line is optically thick. In general, $p_{ij}\ll 1$ for the Lyman lines and $p_{ij}=1$ for all other line transitions. The Sobolev escape probability is included in direct astrophysical equations (see, for example, Eq. (25) \cite{Seager2}) of radiation transfer.

Following section 2.3.3 of Ref. \cite{Seager2} the Sobolev escape probability can be presented in the form:
\begin{eqnarray}
\label{6}
p_{ij}=\frac{1-exp(-\tau_S)}{\tau_S}\, ,
\end{eqnarray}
where $\tau_S$ is the Sobolev optical depth. The optical depth is a measure of the extinction coefficient or absorptivity up to a specific 'depth'. In other words the optical depth expresses the quantity of light removed from a beam by scattering or absorption during its path through a medium. The Sobolev optical depth can be defined as
\begin{eqnarray}
\label{7}
\tau_S=\frac{\lambda_{ij}\tilde{k}}{|v'|}\, ,
\end{eqnarray}
where $\tilde{k}$ is the integrated line absorption coefficient and $\lambda_{ij}$ is the central line wavelength. The monochromatic absorption coefficient or opacity is $k=\tilde{k}\phi(\nu_{ij})$ ($\nu_{ij}$ is the frequency for a given line transition and $\phi(\nu_{ij})$ is the normalized line profile), $v'$ is the velocity gradient which is given by the Hubble expansion rate $H(z)$. 

The absorption coefficient depends strongly on the external conditions and requires the particular consideration for the each case. In presence of an external field the opacity can be related to 
the imaginary part of the density-matrix elements $\rho_{ij}$ as follows:
\begin{eqnarray}
\label{8}
k=\frac{N d_{ij}^2\omega_{ij}}{2\varepsilon_0\Omega_{ij}}Im\big\{\rho_{ij}\big\}\, ,
\end{eqnarray}
where $\varepsilon_0$ is the permittivity of the vacuum and $N$ is the number of atoms.

Using expression (\ref{5}) for the definition of the imaginary part of $\rho_{21}$, we obtain
\begin{eqnarray}
\label{9}
Im\big\{\rho_{21}\big\}=\frac{\gamma_{21}\Omega_{\alpha}/2}{\Delta^2_{\alpha}+\gamma_{21}^2}\left[1+f(\Omega_{\alpha}^2,\Omega_{\beta}^2,\Delta_{\alpha},\Delta_{\beta})\right] 
\end{eqnarray}
together with the dimensionless function 
\begin{eqnarray}
\label{9a}
f(\Omega_{\alpha}^2,\Omega_{\beta}^2,\Delta_{\alpha},\Delta_{\beta})=\frac{\Delta_{\alpha}^2-\gamma_{21}^2}{\Delta_{\alpha}^2+\gamma_{21}^2}\left[\frac{\Omega_{\beta}^2}{4\gamma_{21}\gamma_{31}}-\frac{\Omega_{\alpha}^2}{4\Gamma_2\gamma_{21}}\right]+
\nonumber
\\
+\frac{\left(\gamma_{21}^4\gamma_{32}-\gamma_{32}\Delta_{\alpha}^4-2\gamma_{21}^3\gamma_{32}^2+2\gamma_{21}\Delta_{\alpha}^2(3\gamma_{32}^2+\Delta_{\beta}(3\Delta_{\beta}-\Delta_{\alpha}))\right)\Omega_{\alpha}^2\Omega_{\beta}^2}{16\Gamma_2\gamma_{21}\gamma_{31}(\Delta_{\alpha}^2+\gamma_{21}^2)^2(\Delta_{\beta}^2+\gamma_{32}^2)}+
\\
\nonumber
+
\frac{\left(\gamma_{21}^2\gamma_{32}-\gamma_{32}\Delta_{\alpha}^2-2\gamma_{21}\Delta_{\alpha}\Delta_{\beta}\right)\Omega_{\alpha}^2\Omega_{\beta}^2}{16\gamma_{21}\gamma_{31}^2(\Delta_{\alpha}^2+\gamma_{21}^2)(\Delta_{\beta}^2+\gamma_{32}^2)}+...
\end{eqnarray}

To get informations about the physical meaning of the function $f$, i.e. about the absorption and
subsequent emission processes,  one can proceed 
as earlier for the density-matrix element $\rho_{21}$.
For the definition of the integrated line absorption coefficient from Eqs. (6)-(8) the line profile is separated out. It is assumed that the line profile appearing in Eq. (7) corresponds to the monochromatic absorption coefficient, see Eq. (31) in \cite{Seager2}. Here the function $f$ depends on the fixed parameters $\Delta_{\alpha}$ and $\Delta_{\beta}$, although the common factor represents the Lorentz line profile, where $\Delta_{\alpha}=\omega_{\alpha}-\omega_{21}$.

Thus the integrated line absorption coefficient can be presented in the form:
\begin{eqnarray}
\label{10}
\tilde{k}_{21}=\frac{\pi d_{21}^2 N\omega_{21}}{4\varepsilon_0}\left[1+f(\Omega_{\alpha}^2,\Omega_{\beta}^2,\Delta_{\alpha},\Delta_{\beta})\right]
\end{eqnarray}
with the line profile $\phi({\nu_{21}})=\gamma_{21}/\left[(\omega_{\alpha}-\omega_{21})^2+\gamma_{21}^2\right]$. In accordance with the theory described in \cite{Seager2} the line profile should be normalized 
within the interval $[0,\infty]$ and the coefficient $\pi$ arises in Eq. (\ref{10}). Thus the Sobolev escape probability extends to the expression 
\begin{eqnarray}
\label{11}
p_{12}=\frac{1-exp\left(-\tau_S[1+f(\Omega_{\alpha}^2,\Omega_{\beta}^2,\Delta_{\alpha},\Delta_{\beta})]\right)}{\tau_S[1+f(\Omega_{\alpha}^2,\Omega_{\beta}^2,\Delta_{\alpha},\Delta_{\beta})]}\, ,
\end{eqnarray}
where $\tau_S$ can be taken in the form (\ref{7}), Eqs. (39), (40) in \cite{Seager2}. In principle, expression (\ref{11}) should be employed in further astrophysical evaluations.
We reserve this for forthcoming research and restrict ourselves to the consideration of the function $f(\Omega_{\alpha}^2,\Omega_{\beta}^2,\Delta_{\alpha},\Delta_{\beta})$. The function $f$ depends strongly on the parameters $\Omega_{\alpha}^2,\Omega_{\beta}^2,\Delta_{\alpha},\Delta_{\beta}$. The applicability of the power series expansion Eq. (\ref{5}) with respect to $\Omega_{\alpha}$, $\Omega_{\beta}$ is limited by the field amplitudes and can be found in \cite{Gea-Ban} but beyond in \cite{Wiel}. Estimates for the  field amplitudes can be deduced from the  CMB distribution corresponding to the hydrogen recombination era in earlier universe. 

\section{Numerical results and discussion}

While the universe further expanded and cooled down the electrons and protons tended towards 
formation of hydrogen atoms. The temperature at this era is very well known from the laboratory physics, $T\approx 4500-3000$ K. After "recombination", the photons released were able to travel through the universe relatively undisturbed, and formed the primordial background radiation. 
However, such a photon environment (background) should have influence on the hydrogen atom. 
The field amplitudes for a circular polarized wave can be obtained from the (thermal-averaged)
spectral energy density
\begin{eqnarray}
\label{12}
\frac{c \varepsilon_0 |E|^2}{4\pi}=\frac{2h\nu_{ij}^3\Delta\nu_{ij}}{c^2}\frac{1}{e^{\frac{h\nu_{ij}}{k_B T_e}}-1}\, ,
\end{eqnarray}
where $c$ is the speed of light, $k_B$ is the Boltzmann constant, $h$ is Planck's constant and in further calculations we use $T_e=3000$ K. The right-hand side of the equation above corresponds to the black-body distribution of the CMB, while left-hand side defines the (electrical) energy density.

In order to avoid de-phasing problem we should choose $\Delta\nu_{ij}\sim \Gamma_i$. Hence for the spectral lines $\nu_{21}=\nu_{\alpha}$ ($Ly_{\alpha}$ line) and $\nu_{32}=\nu_{\beta}$ ($H_{\alpha}$ line) we obtain
\begin{eqnarray}
\label{13}
E_{\alpha}\approx 0.000068802\,\, V/m = 1.33799\cdot 10^{-16}\, a.u.,
\nonumber
\\
E_{\beta}\approx 52.8636\,\, V/m = 1.02803\cdot 10^{-10}\, a.u.
\end{eqnarray}

The magnitudes (\ref{13}) of the field are small; we should compare the Rabi frequencies with the corresponding level widths. The one-photon transition rates, which yield the major contributions to level widths, can be easily evaluated and are well known. For the hydrogen atom the dominant one is Ly-$\alpha$ transition rate, $\Gamma_{2p}\sim 10^{-8}$ in atomic units and, therefore, the power series (\ref{5}) is valid. Moreover, the estimates given in Eq. (\ref{13}) 
reveal that we can neglect all terms of the order $\Omega_{\alpha}^2$ and higher.

Further we evaluate the function $f(\Omega_{\alpha}^2,\Omega_{\beta}^2,\Delta_{\alpha},\Delta_{\beta})$ numerically. In case if detuning are defined by $\Delta_{\alpha}\equiv \Gamma_{2p}$ and $\Delta_{\beta}\equiv \Gamma_{2p}+\Gamma_{3s}$ and using Eq. (\ref{2}) we can present the function $f$ in the form:
\begin{eqnarray}
\label{14}
f(\Omega_{\alpha}^2,\Omega_{\beta}^2,\Delta_{\alpha},\Delta_{\beta})\approx 
-1.30494 \cdot 10^{15} \Omega_{\alpha}^2 -2.08127\cdot 10^{30}\Omega_{\alpha}^4
\nonumber
\\
 + 8.67218\cdot 10^{14} \Omega_{\beta}^2 + 5.15573\cdot 10^{29}\Omega_{\beta}^4 + 2.42665\cdot 10^{44}\Omega_{\beta}^6
\\
\nonumber
 + 3.22916\cdot 10^{29}\Omega_{\alpha}^2 - 1.08979\cdot 10^{47}\Omega_{\alpha}^2 \Omega_{\beta}^4 - 6.18661\cdot 10^{44}\Omega_{\alpha}^4 \Omega_{\beta}^2
\\
\nonumber 
 - 1.63588\cdot 10^{62} \Omega_{\alpha}^4 \Omega_{\beta}^4  + 2.92629\cdot 10^{63}\Omega_{\alpha}^2 \Omega_{\beta}^6 - 4.88545\cdot 10^{78}\Omega_{\alpha}^4 \Omega_{\beta}^6.
\end{eqnarray}
Taking into account the estimates (\ref{13}) we receive
\begin{eqnarray}
\label{15}
\Omega_{\alpha}\approx 1.99343\cdot 10^{-16}\, a.u.,
\nonumber
\\
\nonumber
 \Omega_{\beta}\approx 1.92942\cdot 10^{-10}\, a.u.,
 \\
f\approx 0.0000322844.
\end{eqnarray}
In case of exact resonances from Eq. (\ref{2}) and (\ref{13}) we get
\begin{eqnarray}
\label{16}
\Delta_{\alpha}=\Delta_{\beta}\equiv 0 \, ,
\nonumber
\\
f\approx -0.015814.
\end{eqnarray}

Finally, in case of an exact two-photon resonance, where $\Delta_{\alpha}+\Delta_{\beta}=0$ and together with $\Delta_{\alpha}=\Gamma_{2p}$, the function $f$ takes the value:
\begin{eqnarray}
\label{17}
\Delta_{\alpha}=\Gamma_{2p}=-\Delta_{\beta}\sim 10^{-8}\, a.u.
\nonumber
\\
f\approx 0.00952743.
\end{eqnarray}

Thus the magnitude of the function  $f(\Omega_{\alpha}^2,\Omega_{\beta}^2,\Delta_{\alpha},\Delta_{\beta})$ is of about $1\%$ ($10^{-2}$); in case of an exact one-photon resonances of about $1.5\%$, respectively about $0.95\%$ if the detuning are no-zero but of opposite sign. In quantum optics this effect is well known   when the electromagnetically induced transparency is investigated for different kind of systems  (two-, three- or four level systems with $\Lambda$-, $V$- or $\Xi$-scheme of levels). The results of calculations for different values for detuning are compiled in Table 1 and shown 
graphically in the figures 3 and 4. In astrophysical investigations of the CMB formation processes the Lorentz line profile is used for defining of the absorption coefficient. This line profile represents the dominant term in $Im\big\{\rho_{21}\big\}$ and is separated out in  Eq. (\ref{9}). But in addition the function $f(\Omega_{\alpha}^2,\Omega_{\beta}^2,\Delta_{\alpha},\Delta_{\beta})$ in Eq. (\ref{9}) has to be included in astrophysical evaluation of CMB.

\section{Conclusions}

The aim of our paper is to study the EIT phenomenon and its influence on the CMB formation. The  phenomenon of EIT consists in the investigation of the systems response on the external field. During the recombination era the hydrogen atoms reach their ground state via emission of 
photons from all the spectral lines. After recombination photons were able to travel through universe and formed the CMB. Generated by this way the field should affect the primordial atoms. Thus, we are confronted with the consideration of the phenomena of electromagnetically induced transparancy .

We employed the quantum optical evaluation of the integrated line absorption coefficient that assumes application of photon beams (laser). In astrophysics the photon beam diffusion is 
accounted for via the Sobolev optical depth. Usually, the standard line profile is employed for the solution of the recombination problem and description of the CMB  formation processes and the integrated line absorption coefficient is defined via Einstein coefficients in frames of the Sobolev approximation. But the determination of the integrated line absorption coefficient based on the quantum optical techniques allows one to take into account the influence of the external fields on the atom. Accordingly, we derive the absorption coefficient from the imaginary part of the density matrix element and we deduce the magnitudes of fields from the CMB distribution. 

We are let to the additional function $f$ which depends on external conditions. The values of function $f$ are listed in Table 1 for different values of detuning. The dependence on detuning is depicted in figures 3 and 4. The maximal value of the $f$ was found and amounts to about $1.5\%$ in case of exact resonances (when both de-tunings are equal to zero). In case of exact two-photon resonance, when frequencies of fields are close but differ slightly to the corresponding resonances and the total detuning is equal to zero, we obtained $f\approx 0.95\%$. Figures 3 and 4 also reveal that additional transparency of the medium yields contribution at the level about $1\%$. We expect that the modifications of this magnitude should definitely be relevant in determinations of the key cosmological parameters.

The problem of dephasing appears when 
defining the magnitude of the field. The effect of dephasing leads to the spectral line broadening. To prevent dephasing phenomena we confined the definition of the field by a narrow strip. In this case the width of the corresponding line appears as a natural parameter. In our calculations we used the following relation $\Delta\nu_{ij} = \Delta_{\alpha}(\Delta_{\beta})$. In addition we should note the exponential behavior over temperature for the field amplitudes. With increasing of the temperature $T_e$ larger values for the amplitudes could be obtained, see Eq. (\ref{12}). 
Accordingly, the contribution of the EIT effect will become  more significant 
for higher temperatures.

\begin{center}
Acknowledgments
\end{center}
The work was supported by RFBR (grant No. 08-02-00026 and  No. 11-02-00168-a). G.P. acknowledges the financial support from GSI. D. S. is grateful to the Max Planck Institute for the Physics of Complex Systems for financial support and to the Dresden University of Technology for hospitality. The authors acknowledge valuable discussion with Prof. L. N. Labzowsky and Dr. O. Yu. Andreev.

\begin{table}[h!]
{\citation\, Table 1. The numerical results of the function $f(\Omega_{\alpha}, \Omega_{\beta}, \Delta_{\alpha}, \Delta_{\beta})$ for the different magnitudes of detunings are presented. 
In the first column different values for $f(\Omega_{\alpha}, \Omega_{\beta}, \Delta_{\alpha}, \Delta_{\beta})$ are listed, in the second and third column the detunings and field amlitudes are compiled.}
\begin{center}
\begin{tabular}{c|c|c}
\hline
\hline
\multirow{3}{*}{$f\left(\Omega_{\alpha},\Omega_{\beta},\Delta_{\alpha},\Delta_{\beta}\right)$} & $\Delta_{\alpha} \,\,s^{-1}$ & $|E_{\alpha}|\,\,V/m$ \\
 & $\Delta_{\beta}\,\,s^{-1}$ & $|E_{\beta}|\,\,V/m$ \\
 & $\Delta\nu_{ij}=|\Delta_{\alpha}|(|\Delta_{\beta}|)$ in Eq. (\ref{12})&\\
\hline
\multirow{2}{*}{$6.48635\cdot 10^{-7}$} & $W_{21}=6.26826\cdot 10^8$ & $6.8802\cdot 10^{-5}$ \\
 & $W_{32}=6.31696\cdot 10^6$ & $5.30686$ \\
 \hline

\multirow{2}{*}{$0.0000326097$} & $W_{21}=6.26826\cdot 10^8$ & $6.8802\cdot 10^{-5}$ \\
 & $W_{21}+W_{32}=6.33143 \cdot 10^8$ & $53.1293$ \\

  \hline
\multirow{2}{*}{$0.00952743$} & $W_{21}=6.26826\cdot 10^8$ & $6.8802\cdot 10^{-5}$ \\
 & $-W_{21}=-6.26826\cdot 10^8$ & $52.8636$ \\
 
  \hline
\multirow{2}{*}{$0.00952743$} & $-W_{21}=-6.26826\cdot 10^8$ & $6.8802\cdot 10^{-5}$ \\
 & $W_{21}=6.26826\cdot 10^8$ & $52.8636$ \\
 
   \hline
\multirow{2}{*}{$0.0000324474$} & $-W_{21}=-6.26826\cdot 10^8$ & $6.8802\cdot 10^{-5}$ \\
 & $-W_{21}=-6.26826\cdot 10^8$ & $52.8636$ \\
 
   \hline
\multirow{2}{*}{$0.0000326097$} & $-W_{21}=-6.26826\cdot 10^8$ & $6.8802\cdot 10^{-5}$ \\
 & $-W_{21}-W_{32}=6.33143 \cdot 10^8$ & $53.1293$ \\
 
   \hline
\multirow{2}{*}{$4.06852\cdot 10^{-6}$} & $10 W_{21}=-6.26826\cdot 10^9$ & $2.1757\cdot 10^{-4}$ \\
 & $10(W_{21}+W_{32})=6.33143 \cdot 10^9$ & $168.01$ \\
 
  \hline
  \hline

\end{tabular}
\end{center}
\end{table}

\begin{figure}[h]
\begin{center}
\includegraphics[width=3cm]{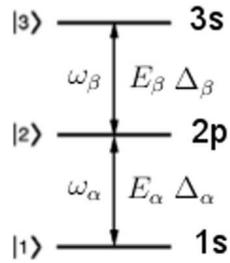}
\end{center}
\caption{\label{Fig1} Scheme of the three-level ladder system under consideration. 
The three levels correspond to the hydrogenic states:  $|1\rangle\rightarrow |1s\rangle$, 
$|2\rangle\rightarrow |2p\rangle$ and $|3\rangle\rightarrow |3s\rangle$, respectively. The frequencies $\omega_{\alpha}$, $\omega_{\beta}$ are the frequencies of external fields 
(which correspond to probe and controlled laser fields in \cite{Wiel}). 
The fields $E_{\alpha}$, $E_{\beta}$ stimulate transitions $1s-2p$ and $2p-3s$ 
(Lyman-$\alpha$ and Balmer-$\alpha$ lines). Possible detunings $\Delta_{\alpha}$ and $\Delta_{\beta}$ for the amplitudes of the fields are also indicated.}
\end{figure}
\begin{figure}[h]
\begin{center}
\includegraphics[width=12cm]{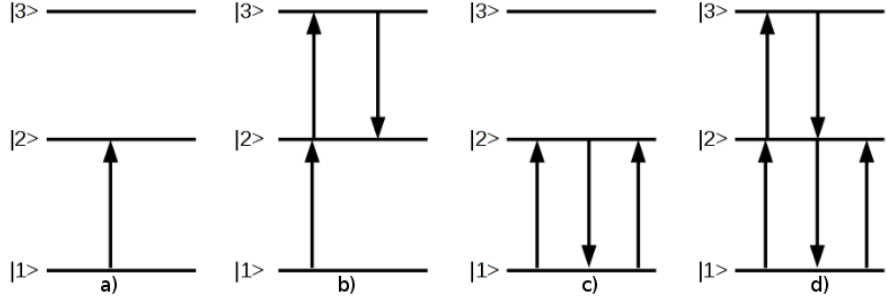}
\end{center}
\caption{\label{Fig2} The transition processes occurring in the three-level ladder system 
corresponding to the different term of Eq. (\ref{5}) are depicted:  
part a) of the figure corresponds to the one-photon absorption processes 
(first term equal to the common factor); part b) and c) represent the second and 
third (product) terms and part d) describes the "interfering-pathways" (fourth term of \ref{5})).} 
\end{figure}
\begin{figure}[h]
\begin{center}
\includegraphics[width=7cm,angle=-90]{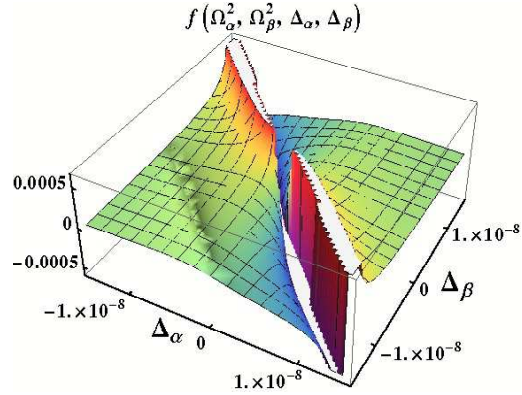}
\end{center}
\caption{\label{Fig3} The dependence of the function $f(\Omega_{\alpha}^2,\Omega_{\beta}^2,\Delta_{\alpha},\Delta_{\beta})$ on the detunings $\Delta_{\alpha}$ and $\Delta_{\beta}$ is depicted for fixed values of the external
fields (\ref{13}), i.e. $\Omega_{\beta}/\gamma_{ij}\ll 1$ and $\Omega_{\alpha}/\gamma_{ij}\ll 1$. The detunings $\Delta_{\alpha}$ and $\Delta_{\beta}$ vary within the range $[-\Gamma_{2p},\Gamma_{2p}]$.}
\end{figure}
\begin{figure}[h]
\begin{center}
\includegraphics[width=7cm,angle=-90]{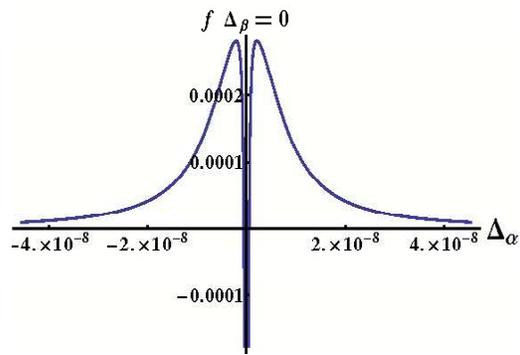}
\end{center}
\caption{\label{Fig8} The section through the 2-dimensional plot of function $f(\Omega_{\alpha}^2,\Omega_{\beta}^2,\Delta_{\alpha},\Delta_{\beta})$ (see \ref{Fig3}) 
at $\Delta_{\beta}=0$ is plotted. The remaining detuning parameter $\Delta_{\alpha}$ 
varies within the range $[-3\Gamma_{2p},3\Gamma_{2p}]$.}
\end{figure}

\end{document}